\documentclass[number,3p,twocolumn]{elsarticle}
\usepackage{amsmath}
\usepackage{amssymb}
\usepackage{graphicx}
\usepackage{fix2col} % added by Hyun Kyung
\usepackage{color} % added by Hyun Kyung
\def\be{\begin{equation}}
\def\bsp{\begin{split}}
\def\a{\alpha}
\def\b{\beta}
\def\g{\gamma}
\def\G{\Gamma}
\def\d{\delta}
\def\D{\Delta}
\def\e{\epsilon}

\def\x{\xi}
\def\p{\pi}
\def\r{\rho}
\def\s{\sigma}

\def\t{\tau}

\def\o{\omega}

\def\i{\int}

\def\bp{{\mathbf p}}
\def\bq{{\mathbf q}}
\def\bP{{\mathbf P}}
\def\bQ{{\mathbf Q}}

\def\bF{{\mathbf F}}

\def\ve{{\mathbf e}}

\def\bk{{\mathbf k}}

\def\bQ{{\mathbf Q}}

\def\cN{{\mathcal N}}

\def\ee#1{\label{#1}\end{equation}}
\def\esq{\end{split}}

\begin{document}
%\preprint{HEP/123-qed}
\title{Brownian motion from molecular dynamics}
\author[Au,Te]{Hyun Kyung Shin}
\ead{julie2020@kaist.ac.kr}
\author[Te]{Changho Kim}
\ead{changhokim@kaist.ac.kr}
\author[Au]{Peter Talkner }
\ead{peter.talkner@physik.uni-augsburg.de}
\author[Te]{Eok Kyun Lee \corref{cor1}}
\ead{eklee@cola.kaist.ac.kr}
\cortext[cor1]{corresponding author}
\address[Au]{Universit\"at Augsburg, Institut f\"ur Physik,
D-86135 Augsburg, Germany}
\address[Te]{Department of Chemistry ,Korea Advanced Institute of Science and Technology, Daejeon
  305-701, Republic of Korea}
\date{\today}
%\maketitle
\begin{abstract}
Brownian motion of single particles with various masses $M$ and diameters $D$ is studied by molecular dynamics simulations. Besides the momentum auto-correlation function of the Brownian particle the memory function and the fluctuating force which enter the generalized Langevin equation of the Brownian particle are determined and their dependence on mass and diameter are investigated for two different fluid densities.  Deviations of the fluctuating force distribution from a Gaussian form are observed for small particle diameters. For heavy particles the deviations of the fluctuating force from the total force acting on the Brownian particle decrease linearly with the mass ratio $m/M$ where $m$ denotes the mass of a fluid particle.
\end{abstract}
\maketitle

\section{Introduction}\label{I}
Brownian motion has a long history probably going back to the end of the eighteenth century when Jan Ingen-Housz\footnote{J. Ingen-Housz was a dutch scholar and famous physician of his time who vaccinated the royal families of king George III and of empress Maria Theresia against smallpox. He was the first to recognize the importance of chlorophyll, the green matter (gr\"une Materie), as he called it, for the production of oxygen by plants.} mentioned the observation of small particles immersed in a fluid that are in motion as if they were living beings although not the slightest vital spark was in them \cite{Ih}. Ingen-Housz presumably misinterpreted this first documented observation of Brownian motion as a consequence of a fluid motion caused by the evaporation of the considered fluid droplet\footnote{He writes: Wenn man sich auch begn\"ugen wollte, die Gestalt und Gr\"osse von einigen dieser K\"orper w\"ahrend des kurzen Zeitraums, als das dem Brennpunkte eines Vergr\"osserungsglases ausgesetzte Tr\"opfchen dauert, zu beobachten, so mu{\ss} man dennoch eingestehen, da{\ss} w\"ahrend der ganzen Zeit, als das Tr\"opchen dauert, dessen best\"andige Verd\"unstung nothwendigerweise den ganzen Saft, und folglich auch die darin enthaltenen K\"orper in eine immerw\"ahrende Bewegung versetze, und da{\ss} diese Bewegung betr\"ugen, und in einigen F\"allen gewisse K\"orperchen als lebendige Wesen darstellen k\"onne, die nicht den geringsten Funken des Lebens haben. Um es klar einzusehen, da{\ss} man sich aus Mangel der Aufmerksamkeit in seinem Urtheile hier\-\"uber betr\"ugen k\"onnte, darf man nur in den Brennpunkt eines Mikroskops einen Tropfen Weingeist sammt etwas gesto{\ss}ener Kohle setzen; man wird diese K\"orperchen in einer verwirrten best\"andigen und heftigen Bewegung erblicken, als wenn es Thierchen w\"aren, die sich reissend untereinander fortbewegen. \cite{Ih,H1}. A very free translation into English reads: The study of small particles (literally translated: bodies) immersed in a small droplet under a magnifying glass is hampered by two effects: The observation time is seriously limited by the evaporation of the droplet (it was  Ingen-Housz who later in this paper suggests the use of thin cover slips that not yet existed at his time); moreover, the evaporation leads to a fluid flow within the droplet that also lets constantly move the immersed particles such that they seem to be alive although not the slightest vital spark is in them. This illusion becomes most obvious if one adds under a microscope some ground coal to a drop of alcohol; one will then see particles in a rapid and permanent motion as if they were small animals that violently moved each other.}.  

In the first half of the nineteenth century Robert Brown published his investigations in which he thoroughly demonstrated that the observed motion of small immersed particles is not restricted to living objects but also occurs for inanimate objects as long as they are small enough \cite{RB}. 

The theoretical understanding of Brownian motion by Sutherland \cite{S},  Einstein \cite{E} and Smoluchowski \cite{Sm} opened several completely new vistas \cite{HM}. Brownian motion has provided a cornerstone of the mathematical theory of stochastic processes in general and of diffusion processes in particular \cite{F}. In physics Brownian motion has been a continuous inspiration for the understanding of such diverse phenomena as transport processes in condensed matter \cite{CL}, statistical mechanics of non-equilibrium processes \cite{Z,vK,HT}, activated rate processes \cite{Ch,HTB,PT}, stochastic resonance \cite{GJHM}, Brownian motors \cite{AH,R}, as well as in nanosciences \cite{HMRMP} and biophysics \cite{FK},
to name but a few.

In particular, the description of Brownian motion by Langevin \cite{L} in terms of a Newtonian equation of motion including damping and random forces has provided an extremely powerful tool to model and  also to numerically simulate such processes ever since we have powerful computers at hand. 

Mori suggested microscopic derivations of so-called generalized Langevin equations \cite{M} that also allow for memory effects. The Mori formalism is based on projection operator techniques that can also be used to derive generalized nonlinear Langevin equations \cite{GHT} as well as corresponding generalized master equations\cite{Z1,GTH}. These projection operator techniques generally provide insight into the general structure of the governing mesoscopic equations and their symmetries \cite{GTH2} in a transparent way but give expressions for transport coefficients, memory functions, fluctuating forces and the like, that often are extremely involved and difficult if not impossible to evaluate apart from certain limiting situations \cite{Spohn}. In the particular case of Brownian motion of a heavy particle of mass $M$ interacting with a large number of fluid particles of mass $m$ a Markovian description of the motion of the heavy particle results if the mass ratio $m/M$ tends to zero \cite{LR}.    

On the other hand, {\it Molecular Dynamics} (MD) provides a direct tool to determine the motion of a Brownian particle in a fluid by numerically solving the Hamiltonian equations of motion for the Brownian and fluid particles all of which may interact with each other \cite{Rapaport,BZ}. In this way the friction coefficients were determined for an infinitely massive particle immersed in truncated Lennard-Jones and hard sphere fluids in Refs.~\cite{EZ} and \cite{BHP}, respectively.  The assumption of infinite Brownian particle mass considerably simplifies the analysis because then the velocity of the Brownian particle vanishes and the total force acting on the Brownian particle coincides with the fluctuating force. Therefore the fluctuating force entering the Langevin equation becomes directly accessible in the MD simulation when the Brownian particle mass is infinite. 

In the present paper we study a single Brownian particle of various {\it finite masses} and {\it diameters} immersed in a truncated Lennard-Jones fluid. The range of parameters that we consider here is relevant for diffusion of medium size molecules like fullerenes in liquids. For applications to nanofluidics the influence of the confinement becomes important but is not considered here.    

In the studied parameter regime the fluctuating force in general differs from the total force by the so-called organized time rate of change of the momentum which contains a memory function weighting the influence of the value of the momentum as it was at earlier times. This memory function though cannot directly be determined from the output of the MD simulations which consists of the position and momentum of the Brownian particle, as well as of the total force acting on the Brownian particle, at each integration time step. In fact, the memory kernel is related to the momentum auto-correlation function by a Volterra integral equation of the first kind \cite{Z}. The numerical solution of this type of equations is known to be error-prone. A conversion to a Volterra equation of the second kind can be achieved, for which more stable algorithms exists, but which involves derivatives of the momentum auto-correlation function with respect to time up to second order. For their efficient estimates Berne and Harp \cite{BH} used local polynomial approximations of the momentum auto-correlation function. Here we do without any fitting. We identify the first and second time derivatives of the momentum auto-correlation function with correlation functions of momentum and total force and with the total force auto-correlation function, respectively. Both correlation functions can be directly estimated from the MD simulation data leading to reliable results for the memory function. 

Kneller and Hinsen suggested an alternative method of determining the memory function \cite{KH}. Their method is based on a fit of the momentum auto-correlation function by an auto-regressive model AR($P$) where the order $P$ determines the maximal extent of the memory time. We do not further pursue this method since it requires the fitting of a large number of auxiliary parameters.    

Once the memory kernel is known, the fluctuating force can also be determined and further analyzed. Of particular interest is the questions under which conditions the random force becomes Gaussian and/or Markovian and when it can be approximated by the total force. 

The paper is organized as follows. In Section~\ref{II} the employed microscopic model is specified and the MD simulation outlined. In Section~\ref{III} the generalized Langevin equation is reviewed and the estimation of the respective memory kernel and fluctuating force based on  MD simulation data is described. The momentum auto-correlation function obtained from the MD simulations is discussed in Section~\ref{IV}. The consistency of the resulting memory kernel with the structure of the generalized Langevin equation imposed by the Mori theory is confirmed in Section~\ref{V}. The mass and diameter dependence of the memory kernel and the fluctuating force is investigated in Sections~\ref{VI} and \ref{VII}. The paper ends with concluding remarks.

\section{Microscopic model of 2d Brownian motion and MD simulations}\label{II}
As a simple microscopic model of Brownian motion we consider $N$ soft ``fluid'' particles of mass $m$ and diameter $d$ and a single ``Brownian'' particle of mass $M$ and diameter $D$ moving in a two dimensional quadratic domain of side-length $L$ with periodic boundary conditions. With $m=M$ and $d=D$ this model includes the case of self-diffusion. The typical regime of Brownian motion though is described by $M\gg m$ and $D \gg d$. The fluid particles interact pairwise with each other as well as with the Brownian particle. The Hamiltonian describing the classical motion of this $N+1$ particle system is of the form
\be
\begin{split}
H= &\frac{1}{2 M} \bP^2 +\sum_{i=1}^{N} \frac{1}{2m} \bp_i^2 +\sum_{i=1}^{N}
V_{\s_B}(|\bq_i-\bQ|) \\
&  +\sum_{i>j}  V_{\s_{fl}}(|\bq_i-\bq_j|)\: ,
\end{split}
\ee{H}
where $\bq_i$ and $\bQ$ denote the positions of the fluid and Brownian particles, respectively, and $\bp_i$ and $\bP$ the according momenta. The interaction $V(r)$ is purely repulsive and equally acts between pairs of fluid particles and between the Brownian particle and fluid particles. It is given by a truncated, purely repulsive Lennard-Jones potential of the form
\be
V_{\s}(r) = \left \{ \begin{array}{ll}
\!4 \e\left [ \left( \s/r\right )^{12} - \left( \s/r\right)^{6} \right] +\e  & \text{for } r < 2^{1/6} \s \\
\!0 & \text{for }  r \ge 2^{1/6} \s \:.
\end{array}
\right .
\ee{Vr}
In the MD simulations dimensionless units were used for which the fluid particle assume the diameter $d=\s_{fl}=1$. For Brownian particles of diameter $D$ we chose $\s_B=(D+d)/2$. Hence, $\s_B$ is the contact distance between the Brownian particle with diameter $D$ and a fluid particle with diameter $d$. Masses are given in multiples of the fluid particle mass and the energy unit $\e$ is chosen as difference of the pair potential energy at distances $r=\s_{fl}$ and $r= 2^{1/6}\s_{fl}$, $\e \equiv V_{\s_{fl}}(\s_{fl})-V_{\s_{fl}}(2^{1/6}\s_{fl})$. A consistent unit of time is then given by $\t = d \sqrt{m/\e}$. 

Then the reduced parameters length $r^*$, temperature $T^*$, and time $t^*$ are defined as $r^* = r / d$, $T^* =T k_B / \e$, and $t^* = t / \t$. If $\eta^*$ is defined as the ratio of the area occupied by the fluid particles to the total available area $\eta^*=N\p d^2/(4 L^2)$, the reduced density is defined as $n^* = 4\eta^*/\p$. We have considered a fluid system composed of $N = 10,000$ fluid particles and one Brownian particle. The simulations were performed at two different densities, one at $n^*=0.4$, and the other one at $n^*=0.8$. The temperature was always $T^* =1$.

Initially the fluid particles occupy lattice points of the (111) face of an fcc lattice. To each particle a random two dimensional vector $\bp^0_i = \sqrt{2 m k_B T}\ve_i$ with identically, uniformly distributed unit vectors $\ve_i$ is assigned. By subtracting the average $\bp^0 = N^{-1} \sum_{i=1}^N \bp^0_i$ we generated initial values of the fluid particle momenta $\bp_i = \bp^0_i -\bp^0$ such that the total fluid is at rest. 

The initial position and momentum of the Brownian particle both are zero. The simulations were realized at constant energy using the standard velocity Verlet algorithm \cite{Rapaport} with a time step $h = 10^{-3}\t$ to insure the stability of the total energy to within $10^{-3}\%$.  

We found that it never took more than $5 \times 10^6$ time steps until thermal equilibrium for fluid and Brownian particle was established. From the subsequent $3\times 10^7$ time steps every third one was used for the reconstruction of a generalized Langevin equation as described in the rest of this paper.

\section{Generalized Langevin equation and its estimation from molecular dynamics simulations}\label{III}
\subsection{Mori's generalized Langevin equation}\label{III.1}
The Mori theory provides a framework to determine the equilibrium correlation functions of a set of so-called ``relevant'' or ``macroscopic'' variables as well as the relaxation of their mean values close to equilibrium in terms of generalized Langevin equations. These equations express the time rate of change of the relevant variables as the sum of systematic and random contributions, the latter also  known as the random force. The relevant part is a linear expression in the relevant variables and in general consists in an instantaneous reversible and a retarded contribution. The retarded contribution is determined by a memory kernel that is connected to the auto-correlation function of the random part via a fluctuation dissipation theorem of second kind. The correlation of the random force with the relevant variables vanishes. Higher order correlation functions of the fluctuating force and consequently higher than second moments of the relevant variables remain unspecified within Mori theory which hence does not provide a complete characterization of the stochastic process of a given set of relevant variables.

In the present study of free Brownian motion we choose the momentum $\bP$ of the Brownian particle as the relevant variable. Due to the absence of an external potential there is no instantaneous contribution to the systematic part of the momentum time rate of change. Hence, the generalized Langevin equation takes the form
\be
\dot{\bP}(t) = - \i_0^t ds \bk(t-s) \bP(s) + \bF^+(t),
\ee{LEQ}
where the mean value of the fluctuating force vanishes, i.e.
\be
\langle \bF^+(t)\rangle =0\: .
\ee{FP}
Moreover, the  correlation functions of the fluctuating force components
$F_\a(t)$ are related to the components of the memory kernel $k_{\a,\b}(t)$
by the fluctuation dissipation theorem
\be
\langle F^+_\a(t) F^+_\b(s) \rangle = \sum_\g \langle P_\a P_\g \rangle
k_{\g,\b}(t-s)\:. 
\ee{FDT}
Due to the isotropy of the present microscopic model specified in the previous section all non-diagonal components vanish and the diagonal components agree with each other
\be
\begin{split}
\langle F^+_\a(t) F^+_\b(s) \rangle & = \d_{\a,\b}\langle F^+(t) F^+(s) \rangle\:,\\
\langle P_\a P_\b \rangle &= \d_{\a,\b} \langle P^2 \rangle\:, \\
 k_{\a,\b}(t-s)&= \d_{\a,\b} k(t-s)\:.
\end{split}
\ee{iso}
Here $F^+(t)$ and $P$ denote either of the $x$ or $y$ components of the fluctuating force and momentum, respectively. Consequently, both components satisfy the same scalar generalized Langevin equations reading
\be
%\begin{split}
\dot{P}(t) =  -\!\i_0^t ds\: k(t-s) P(s) + F^+(t)\: ,
%\end{split}
\ee{sLE}
with $F^+(t)$ obeying the scalar fluctuation dissipation theorem
\be
\langle F^+(t) F^+(s)\rangle  = \langle P^2 \rangle\: k(t-s)\:.
\ee{sfdt}
The microscopic derivation of the generalized Langevin equation (\ref{sLE}) and the corresponding microscopic expressions of the fluctuating force and the memory kernel are based on an identity for the microscopic time evolution operator $e^{-\mathcal{L} t }$ reading
\be
\begin{split}
&e^{-\mathcal{L} t} =  e^{-\mathcal{L}t} \mathcal{P} +(1- \mathcal{P})
e^{-(1-\mathcal{P}) \mathcal{L} t} (1-\mathcal{P})\\
&- \i_0^t ds\: e^{-\mathcal{L}(t-s)} \mathcal{P}
  \mathcal{L}(1-\mathcal{P}) e^{-(1-\mathcal{P})\mathcal{L} s}
  (1-\mathcal{P})\:,
\end{split}
\ee{oi}
where $\mathcal{L}$ denotes the Liouville operator governing the time evolution of the microscopic phase space probability density $\r$
\be
\begin{split}
\frac{\partial \r}{\partial t} &= \mathcal{L} \r\\
& \equiv \left \{H,\r \right \}\:.
\end{split}
\ee{Le}
The operator $\mathcal{P}$ provides an orthogonal projection of phase space functions $f$ onto the linear subspace of relevant variables which are the $x$ and $y$ components of the Brownian particle momentum in the case of free Brownian motion. It is defined as
\be
\mathcal{P} f = \sum_\a P_\a \left ( P_\a,f \right )/\left ( P_\a,P_\a \right )\:,
\ee{P}
where
\be
\left (g,f \right ) = \i d\G g f e^{-H/k_B T}/ \i d\G e^{-H/k_B T}
\ee{Msp}
denotes the Mori scalar product of phase space functions $g$ and $f$ with respect to the Maxwell-Boltzmann distribution  $e^{-H/k_B T}/ \i d\G e^{-H/k_B T}$ describing thermal equilibrium of fluid and Brownian particle at the temperature $T$. Here the integral extends over the phase space with the volume element $d\G = dP dQ \prod_i^N dp_i dq_i$. Hence, the Mori scalar product
of two phase space functions agrees with the thermal expectation value of the product of these functions, i.e.
\be
\left ( f,g \right ) = \langle f g \rangle \:.
\ee{Meq}

Applying the identity (\ref{oi}) to the time rate of change of either component of the momentum given by $\dot{P}(t) = - e^{-\mathcal{L}t} \mathcal{L} P$ one obtains the generalized Langevin equation (\ref{sLE}) with the following microscopic expressions for the fluctuating force and for the memory kernel:
\begin{align}
\label{fF}
F^+(t)&=(1-\mathcal{P})\exp \left
  \{(1-\mathcal{P})\mathcal{L}t
\right \}\dot{P}\:,\\
k(t)&= \langle \dot{P} e^{-(1-P)\mathcal{L}s} \dot{P} \rangle\:,
\label{k}
\end{align}
where $\dot{P}=- \mathcal{L} P$.

By a scalar multiplication of both sides of eq. (\ref{sLE}) with
$P(0)$ one obtains
the equation of motion for the momentum auto-correlation function
\be
C(t) = \langle P(t) P \rangle
\ee{C}
reading
\be
\dot{C}(t) = - \i_0^t ds\:k(t-s) C(s)\:.
\ee{dC}
Thereby one uses the fact that the fluctuating force $F^+(t)$ and the momentum
$P(0)$ are orthogonal, i.e.
\be
\left (F^+(t),P(0) \right ) =0\:,
\ee{fFP}
as follows from eq. (\ref{fF}).

\subsection{ Memory kernel and fluctuating force from MD simulations}\label{III.2}
From an MD simulation of a system of fluid particles interacting with a single Brownian particle as described in Sect.~\ref{II}, the knowledge of the instantaneous positions and momenta of all particles allows one to directly obtain the momentum of the Brownian particle as well as the total force acting on the Brownian particle. However, the separation of the total force into an organized and a random contribution as it is presented in the generalized Langevin equation (\ref{sLE}) cannot be inferred from the instantaneous microscopic state of the total system. In order to achieve this separation one first estimates the stationary momentum auto-correlation function $C(t)$, see eq. (\ref{C}) and then uses the equation of motion of the momentum auto-correlation function (\ref{dC}) in order to determine the memory kernel.

At first sight it is tempting to employ a Laplace transformation which changes the convolution of the memory kernel and the momentum auto-correlation function into a product and immediately leads to an explicit expression for the Laplace transformed kernel. However, the extreme sensitivity of the inverse Laplace transform to numerical errors renders a reliable determination of the memory kernel in the time domain practically impossible.

Once the momentum auto-correlation function is known, eq. (\ref{dC}) represents a Volterra integral equation of first kind for the memory kernel. A discretization of this equation in principle can be solved quite effectively because it involves the inversion of a T\"oplitz matrix. However, also this method is plagued by numerical inaccuracies.

More stable algorithms exist for Volterra equations of second kind. Any Volterra equation of first kind can be transformed to a Volterra equation of second kind by differentiation with respect to the independent variable. In the present case of eq. (\ref{dC}) this yields
\be
\ddot{C}(t) =-C(0)\: k(t) - \i_0^t ds\:\dot{C}(t-s) k(s)\:.
\ee{ddC}
Here, however, the numerical differentiation of the momentum auto-correlation function, which is not analytically known, may introduce large numerical errors. Berne and Harp \cite{BH} attempted to keep these errors under control by applying fourth and sixth order polynomial approximations  of the velocity auto-correlation for small and large times, respectively \cite{BH}. We  here choose a different strategy avoiding any numerical differentiation. The first and second derivatives of the momentum auto-correlation function can be expressed as the total-force-momentum correlation function and the total-force auto-correlation function, respectively. Strictly speaking, we have
\begin{align}
\label{FtP}
\dot{C}(t)& = \langle F(t) P \rangle\:,\\
\ddot{C}(t)& = -\langle F(t) F \rangle\: .
\label{FtF}
\end{align}
These correlation functions with the total force can be directly estimated from the MD simulation and therefore do not introduce additional errors. Hence, we determined the memory kernel from a numerical solution the integral equation
\be
\langle F(t) F \rangle = C(0)\:k(t) + \i_0^t ds\:\langle F(t-s) P \rangle\: k(s)\; ,
\ee{V2}
which follows from eqs.(\ref{ddC}),(\ref{FtP}) and (\ref{FtF}). The numerical scheme to solve eq.(\ref{V2}) is rather straightforward. The discretization of eq.(\ref{V2}) reads
\be
\begin{split}
&\langle F(i\D t) F \rangle\\
& =  C(0)\:k(i\D t)  + \D t \sum_{j=0}^{i} \o_j \:\langle F(i\D t-j\D t) P \rangle\: k(j\D t)\; ,
\end{split}
\ee{V2_discrete}
where $\o_j =1/2$ for $j=0,i$ and $\o_j=1$ otherwise are a weight factors for the integration. Then, $k(t)$ at every $t=i\D t$ can be obtained iteratively as
\be
\begin{split}
& k(i\D t) = \left\lbrace C(0) + \D t \o_i \langle FP\rangle \right\rbrace^{-1} \times \\
& \left\lbrace \langle F(i\D t)F\rangle - \D t \sum_{j=0}^{i-1} \o_j \langle F((i-j)\D t)P\rangle\:k(j\D t) \right\rbrace,
\end{split}
\ee{kt}
with initial condition $k(0)=\langle F^2\rangle / C(0)$. Such obtained memory kernels for different mass, diameter and density values are displayed in the Figs.~\ref{f4} and \ref{f5} and further discussed in Section~\ref{VI}.  

Once the memory kernel is known, the fluctuating force can be calculated as the difference of the total force that is taken from the MD simulation and the systematic part of the force, i.e we have
\be
F^+(t) = F(t) + \i_0^t ds\:k(t-s) P(s)\:,
\ee{FpF}
where the values of the momenta at the times $s$ prior to $t$ are also taken from the MD simulation. Probability densities of the fluctuating force estimated from histograms for different masses and diameters of the Brownian particle and different densities are displayed in Fig.~\ref{f6} and further discussed in Section~\ref{VII}.  

\subsection{Initialization and stationarity}\label{III.3}
The Mori equation (\ref{sLE}) allows for relaxations of the momentum as long as the initial momentum is sufficiently small such that it can be described within linear response theory. This means that the initial probability distribution of the fluid plus Brownian particle system must be of the form
\be
\begin{split}
&\r_0(\bQ,\bP,\bq_,\bp)= Z^{-1}(\bP_0) e^{-\left [H -\bP \cdot
    \bP_0/M \right]/k_B T}\\
& \approx Z^{-1}(0) e^{- H/k_B T}\left ( 1+\frac{1}{M} \bP_0 \cdot \bP
\right )\:,
\end{split}
\ee{ic}
where $\bP_0$ denotes the averaged initial momentum and $Z(\bP_0)$ the partition function that is defined as
\be
Z(\bP_0)= \i d\G \: e^{-\left [H -\bP \cdot \bP_0/M \right]/k_B}.
\ee{ZP0}
This corresponds to a situation in which the bath is in thermal equilibrium in the presence of a Brownian particle moving on average with momentum $\bP_0$.

In the MD simulations it though is more practical to start with initial conditions in which neither the bath nor the Brownian particle are at equilibrium and then let run the simulation long enough until equilibrium has established. We describe the resulting stationary process by a generalized Langevin equation in which the time of the initial preparation is shifted to the infinitely remote past, i.e. we do not make reference to any particular initial condition and consider
\be
\dot{P}(t) = -\i_{-\infty}^t ds \: k(t-s) P(s) + F^+(t)\:.
\ee{ssLE}

In order to be consistent with the Mori equation (\ref{sLE}) we require the relation (\ref{dC}) between the memory kernel $k(t)$ and the momentum auto-correlation to hold. Then the fluctuating force and the momentum are correlated in the following way
\be
\langle F^+(t) P \rangle = \i_0^\infty ds \: k(t+s) C(s)\: .
\ee{FpP}
In contrast, in the framework of the Mori equation (\ref{sLE}) the initial momentum and the fluctuating force at positive times are uncorrelated as it follows from eq. (\ref{fFP}) because of the initial preparation (\ref{ic}). For a formal proof of eq. (\ref{FpP}) see the appendix~\ref{A}. Moreover, one can show that the stationary Langevin equation in conjunction with the relation (\ref{FpP}) implies the fluctuation dissipation theorem (\ref{FDT}). A proof is also given in appendix~\ref{A}. Vice versa, if one imposes the fluctuation dissipation theorem and the relation (\ref{FpP}) then the equation of motion for the momentum auto-correlation function, eq. (\ref{dC}), is recovered, as proved in Appendix~\ref{A}. Therefore, the Mori equation (\ref{sLE}) and the generalized Langevin equation (\ref{ssLE}) provide equivalent descriptions of the equilibrium properties of the Brownian particle.

\section{Momentum auto-correlation functions}\label{IV}
\begin{figure}
\includegraphics[width=7cm]{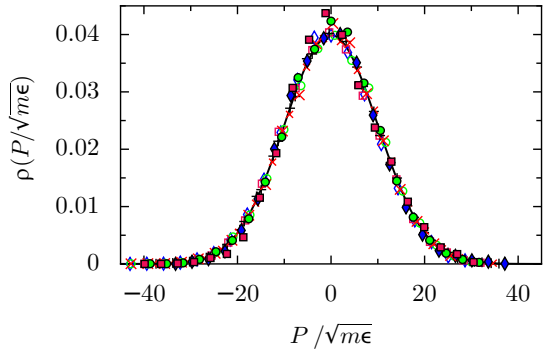}
\caption{(color online) The probability density of a momentum component $\r(P)=\exp\{-P^2/(2 M k_B T)\}/\sqrt{2 \p M k_B T} $ (solid line) for $M=100 m$ is compared to the histograms containing the $x$- and $y$-components of the momentum obtained from MD simulations for Brownian particles with different diameters and two different fluid densities: $D=10d$ (small ($n^* =0.8$) and large ($n^*=0.4$) black plus), $D=7d$ (empty ($n^* =0.8$) and filled ($n^*=0.4$) blue diamond), $D=5d$ (small ($n^* =0.8$) and large ($n^*=0.4$) red cross), $D = 2d$ (open ($n^* =0.8$) and filled ($n^*=0.4$) green circles), and $D=d$  (open ($n^* =0.8$) and filled ($n^*=0.4$) magenta square).
The agreement between theory and simulation is very good. }
\label{fpgauss}
\end{figure}
\begin{figure*}
\includegraphics[width=15cm]{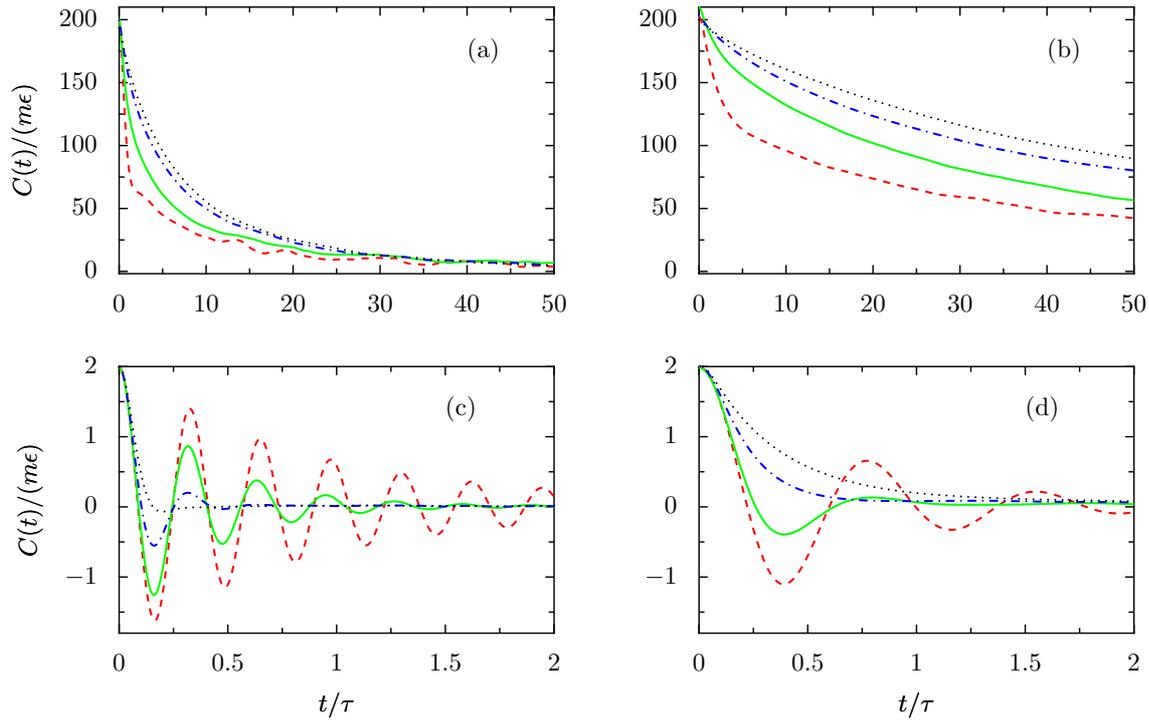}
\caption{(color online) The momentum auto-correlation function $C(t)$ is displayed for Brownian particles of mass $M=100 m$ in panels (a) and (b) and $M=m$ in panels (c) and (d). In panels (a) and (c) the fluid density is $n^*=0.8$, and $n^*=0.4$ for panels (b) and (d).  In each panel different  diameters $D=10 d$ (red, dashed line), $D=5 d$ (green, solid line), $D=2 d$ (blue, dash-dotted line) and $D=d$ (black, dotted line) are shown. All cases shown correspond to the temperature $T^*=1$.  The larger density causes more frequent collisions of the Brownian particle with fluid particles and consequently a faster decay of the momentum auto-correlation function. The pronounced oscillations of the auto-correlation function of large, light particles indicates the presence of an almost empty cavity in which the particle moves back and forth, see panels (c) and (d). The period of oscillations is independent of the particle diameter. For smaller particles the damping of the oscillations increases. The oscillation period is larger for the low density fluid with $n^*=0.4$.      
 }
\label{f1}
\end{figure*}
\begin{figure*}
\includegraphics[width=15cm]{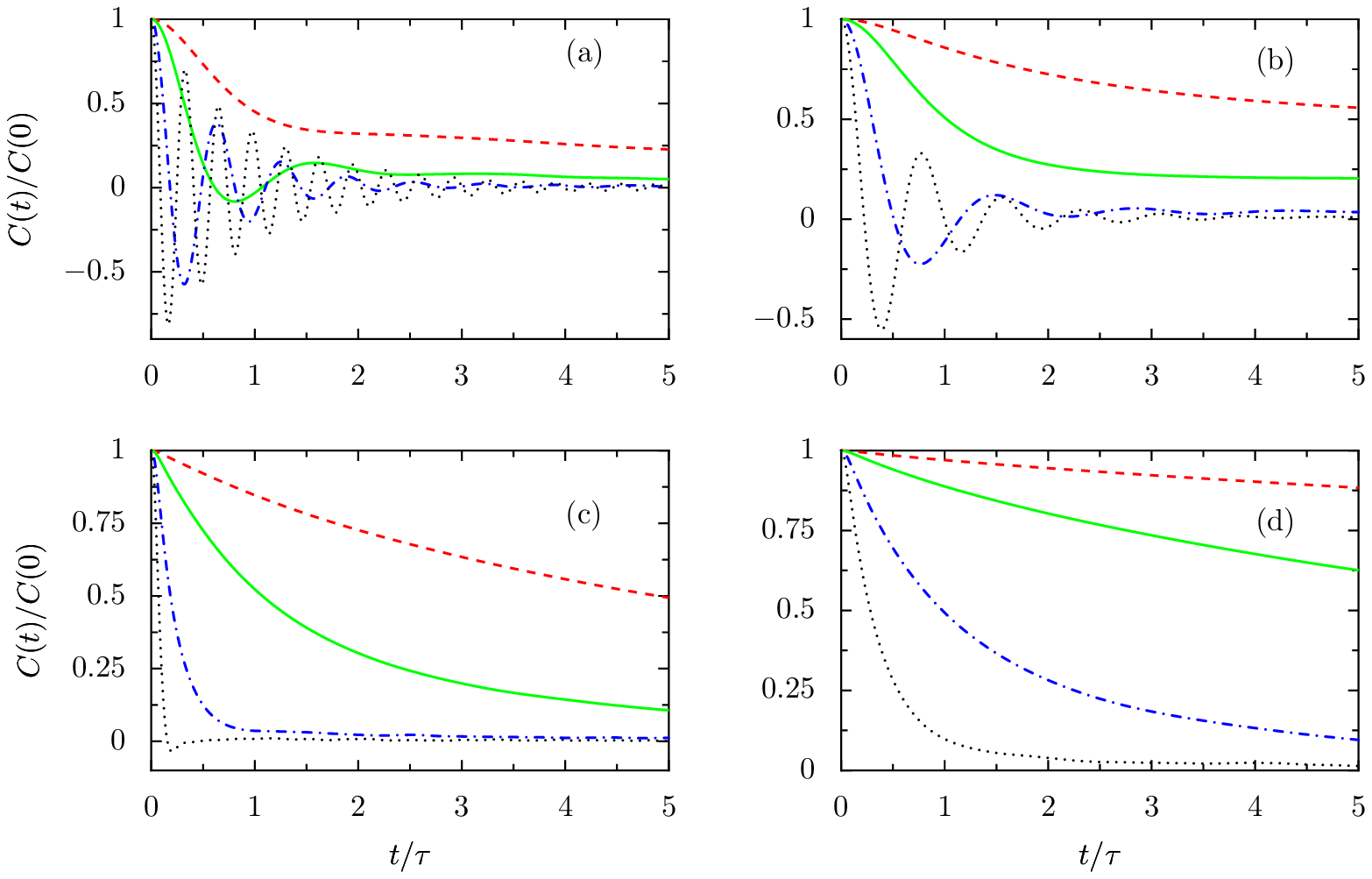}
\caption{(color online) The mass dependence of the normalized momentum auto-correlation function $C(t)/C(0)$ is displayed  for Brownian particles with diameter $D=10 d$  in panels (a) and (b) and for
diameter $D=d$ in panels (c) and (d). Panels (a) and (c) are for a high fluid density $n^*=0.8$ and panels (b) and (d) for the lower density $n^*=0.4$. In all panels the masses are $M=100m$ (red, dashed line), $M = 25 m$ (green, solid line), $M=4m$ (blue, dash-dotted line) and $M=m$ (black, dotted line). 
The pronounced oscillations for large light particles disappear with increasing mass, see (a) and (b), and are absent for small particles of radius $D=d$ at any mass, see (c) and (d). Lower density leads to a slower decay and in case of oscillations to a longer period.
 }
\label{f2}
\end{figure*}
\begin{figure}
\includegraphics[width=7cm]{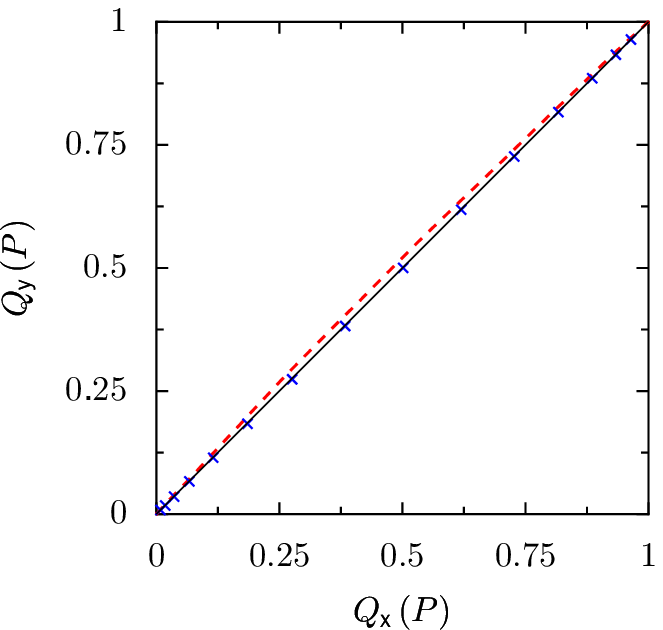}
\caption{(color online) The QQ-plot of the $y$ and $x$ momentum components displays a slight deviation between the cumulative distribution functions $Q_\a(P) = \text{Prob}(P_\a<P)$ with $\a = x,y$ for a heavy Brownian particle with mass $M=100m$ (red, dashed line) in contrast to the case of a light particle with $M=m$ (blue crosses). In both cases particles have the same diameter $D=10 d$ and move in a fluid of density $n^*=0.8$. The ideal relation $Q_y =Q_x$ indicating identical distributions is displayed by the thin black line. The seeming anisotropy of the momenta is of statistical nature due to the long-lived momentum auto-correlation function of heavy Brownian particles. 
 }
\label{fpxpy}
\end{figure}

After a transient period, the Brownian particle has equilibrated. Accordingly, the momentum distribution of the Brown\-ian particle becomes Maxwellian. In particular, the momentum distributions are independent of the size of the Brownian particles and of the fluid density; moreover, the  widths of the distributions conform with the equipartition law, $\langle P_\a^2 \rangle =M k_B T$ for each component. These findings are illustrated in Fig.~\ref{fpgauss} for Brownian particles of mass $M =100 m$ and different diameters and densities. The simulation results for other masses of the Brownian particle also conform with theory but are not shown. 

The stationary momentum auto-correlation function $C(t)= \langle P(t) P(0) \rangle$ is estimated as time average of the stationary part of the simulated time series of momenta, i.e.
\be
C(t) = \frac{1}{\cN(t)} \sum_{j=1}^{\cN(t)} P(t_0+j\D t + t) P(t_0 + j\D t)
\ee{Ct}
where $t_0$ is the time after which equilibrium has established. The upper limit of the sum is given by $\cN(t)=\cN-t/\D t$ where $\cN$ is the total number of momenta. The correlation function was estimated for every third time step $\D t = 3h$. Results are displayed for Brownian particles with mass equaling the fluid mass and different diameters in Figs.~\ref{f1}(c) and \ref{f1}(d) and for  heavy Brownian particles in Figs.~\ref{f1}(a) and \ref{f1}(b). The dependence of the normalized momentum auto-correlation function $C(t)/C(0)$ on mass is illustrated for small and large Brownian particles in Fig.~\ref{f2}, respectively. The momentum auto-correlation function of light and large  particles is characterized by decaying oscillations whereas the decay becomes monotonic in the case of heavy or small particles. For the lower density the period of oscillations is larger and the speed of decay is decreased due to a larger mean free path and mean time  between collisions. The oscillations observed for large light Brownian  particles indicate the presence of a cavity surrounded by fluid particles within which the Brownian particle moves back and forth. 

An estimate of the large time behavior of the momentum auto-correlation function is a notoriously difficult problem \cite{ltt}. These long-lived correlations though deteriorate the estimate of the momentum statistics as we will discuss now. By comparing the distributions of the $x$ and $y$ components of the momentum one finds slight deviations, see Fig.~\ref{fpxpy}. These deviations can mainly be attributed to differences of the estimated mean values $\bar{P}_\a= \sum_{i=1}^\cN P_\a(t_0 + i \D t)/\cN$, $\a=x,y$. The time averages $\bar{P}_\a$ themselves are random quantities having the ensemble average $\langle \bar{P}_\a \rangle =0$ and the variance $\langle \bar{P}_\a^2 \rangle = (C(0) + 2\sum_{i=1}^\cN (\cN-i)C(i\D t))/\cN^2 \approx 2\int_0^{\mathcal{T}} dt(\mathcal{T}-t) C(t)/\mathcal{T}^2$. The approximation by an integral is valid in the case of oversampled data, i.e. if $C(t)$ changes only little with the sampling time $\D t$. In any case, the variance of the momentum time average  increases when the correlations extend over large times, explaining the slight, seeming anisotropy of the momentum statistics for heavy Brownian particles. For light Brownian particles the momentum auto-correlation decays much faster, (cf. Figs.~\ref{f1} and \ref{f2}), and therefore the estimated momentum distribution conforms much better with the expected isotropy, see Fig~\ref{fpxpy}. 

A better estimate of the statistics of the momenta can be obtained if it is based on several shorter independent trajectories instead of a single long one. We confirmed this by comparing estimates from ten trajectories consisting of $10^6$ sampled momenta with a single one with $10^7$ sampled momenta. The former indeed led to better convergence. However, since we here are mainly interested in the memory kernel and the statistical properties of the fluctuating forces, these quantities are essentially determined by the time correlations of the momenta based on single long trajectories.        
    
\section{Two consistency checks}\label{V}
\begin{figure*}%[tp]
\includegraphics[width=15cm]{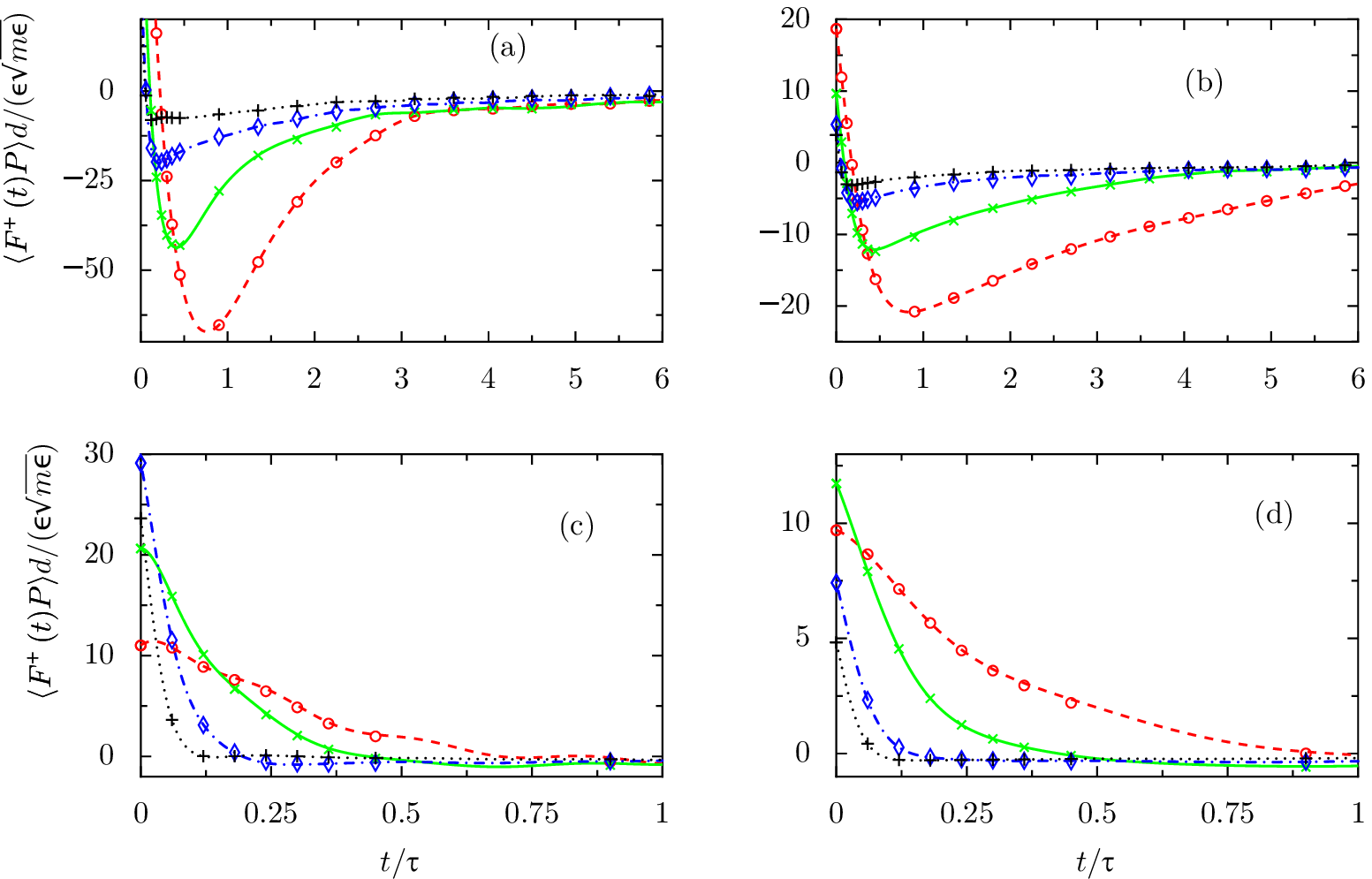}
\caption{(color online) The stationary fluctuating force momentum correlation function $\langle F^+(t) P \rangle$ displayed as lines is compared to the integral $\i_0^\infty ds k(t+s) C(s) $ (symbols) for heavy Brownian particles with Mass $M=100m$ in panels (a) and (b) and light Brownian particles, $M=m$ in panels (c) and (d). Panels (a) and (c) refer to the density $n^*=0.8$ and panels (b) and (d) to the lower density $n^*=0.4$. All panels contain results for different diameters $D=10 d$ (red, dashed line; red circle), $D=5 d$ (green, solid line; green cross), $D=2 d$ (blue dash-dotted line; blue diamond) and $D=d$ (black, dotted line; black plus). In all cases the agreement is perfect in accordance with eq.~(\ref{FpP}). }
\label{f3}
\end{figure*}
\begin{figure*}%[tp]
\includegraphics[width=15cm]{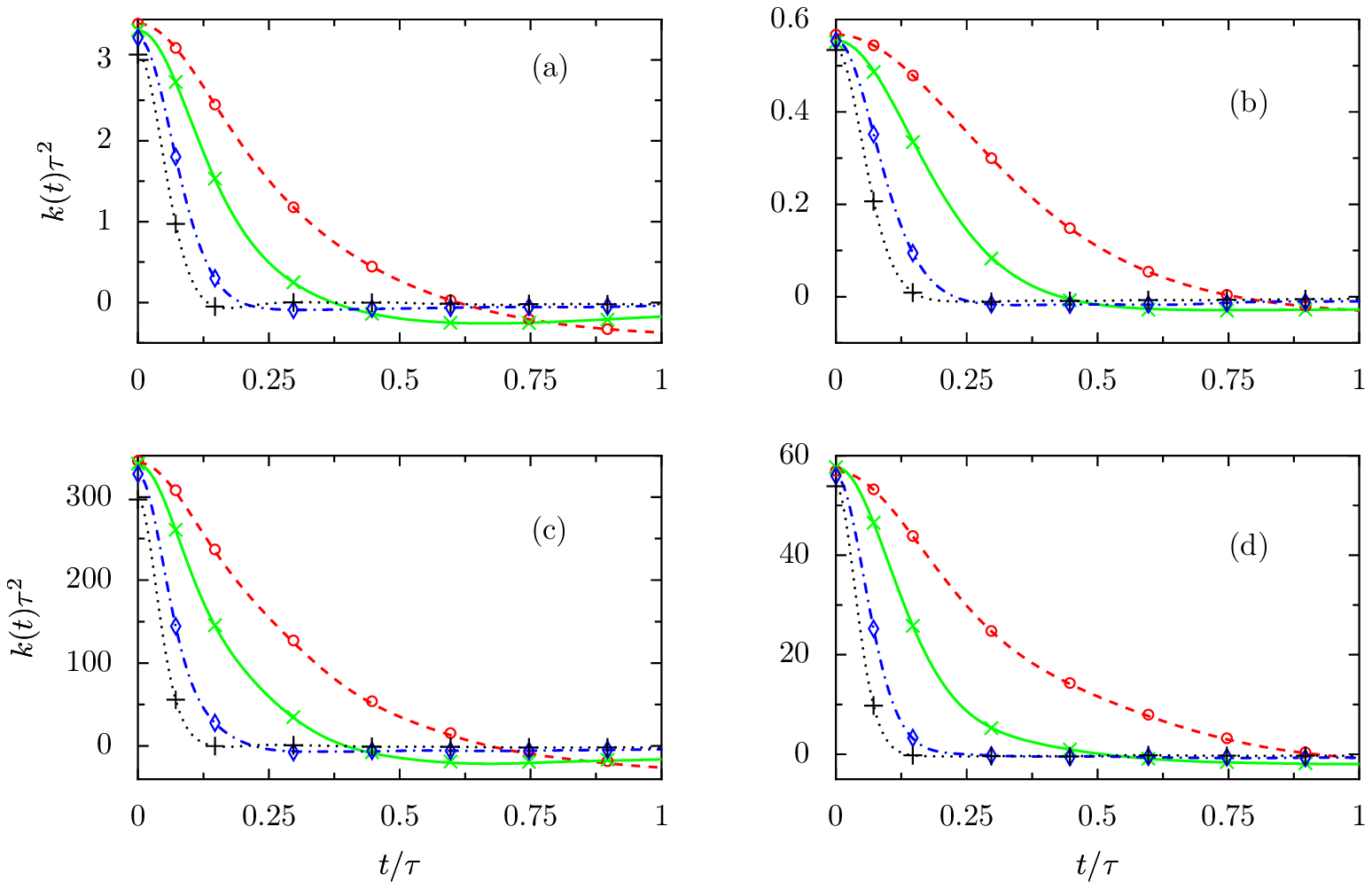}
\caption{(color online) The memory function $k(t)$ displayed as lines is compared to the normalized fluctuating force auto-correlation function $\langle F^+(t) F^+ \rangle / C(0)$ shown by symbols for heavy Brownian particles with Mass $M=100m$ in panels (a) and (b) and light Brownian particles, $M=m$ in panels (c) and (d). Panels (a) and (c) refer to the density $n^*=0.8$ and panels (b) and (d) to the lower density $n^*=0.4$. All panels contain results for different diameters $D=10 d$ (red, dashed line; red circle), $D=5 d$ (green, solid line; green cross), $D=2 d$ (blue dash-dotted line; blue diamond) and $D=d$ (black, dotted line; black plus). In all cases the agreement is perfect in accordance with the fluctuation dissipation theorem (\ref{sfdt}).}
\label{f4}
\end{figure*}

Before discussing the memory kernel and the statistical and temporal properties of the fluctuating force in more detail we consider the consistency of the estimates of these  quantities obtained from MD simulations by checking the relations (\ref{FpP}) and (\ref{sfdt}). 

The stationary fluctuating force-momentum correlation function $\langle F^+(t)P \rangle$ entering eq. (\ref{FpP}) was estimated from MD simulation results for Brown\-ian particles and compared with the integral of the product of the memory kernel shifted in time and the momentum auto-correlation function. Also these functions were determined from the MD simulation data. Fig.~\ref{f3} displays a perfect agreement of the fluctuating force-momentum auto-correlation function and this integral for several masses and diameters of the Brownian particle as well as for two densities of the fluid. 

We also find perfect agreement of the fluctuating force auto-correlations function normalized by $C(0)= \langle P^2\rangle $ with the memory kernels for the same set of Brownian particle and fluid parameters as above, see Fig.~\ref{f4}. Hence the simulated MD data of the Brownian particle momenta are perfectly described by a generalized Langevin equation with the specified memory kernel and fluctuating forces related by the fluctuation dissipation theorem.

\section{Brownian mass and diameter dependence of the memory kernel}\label{VI}
\begin{figure*}%[tp]
\includegraphics[width=15cm]{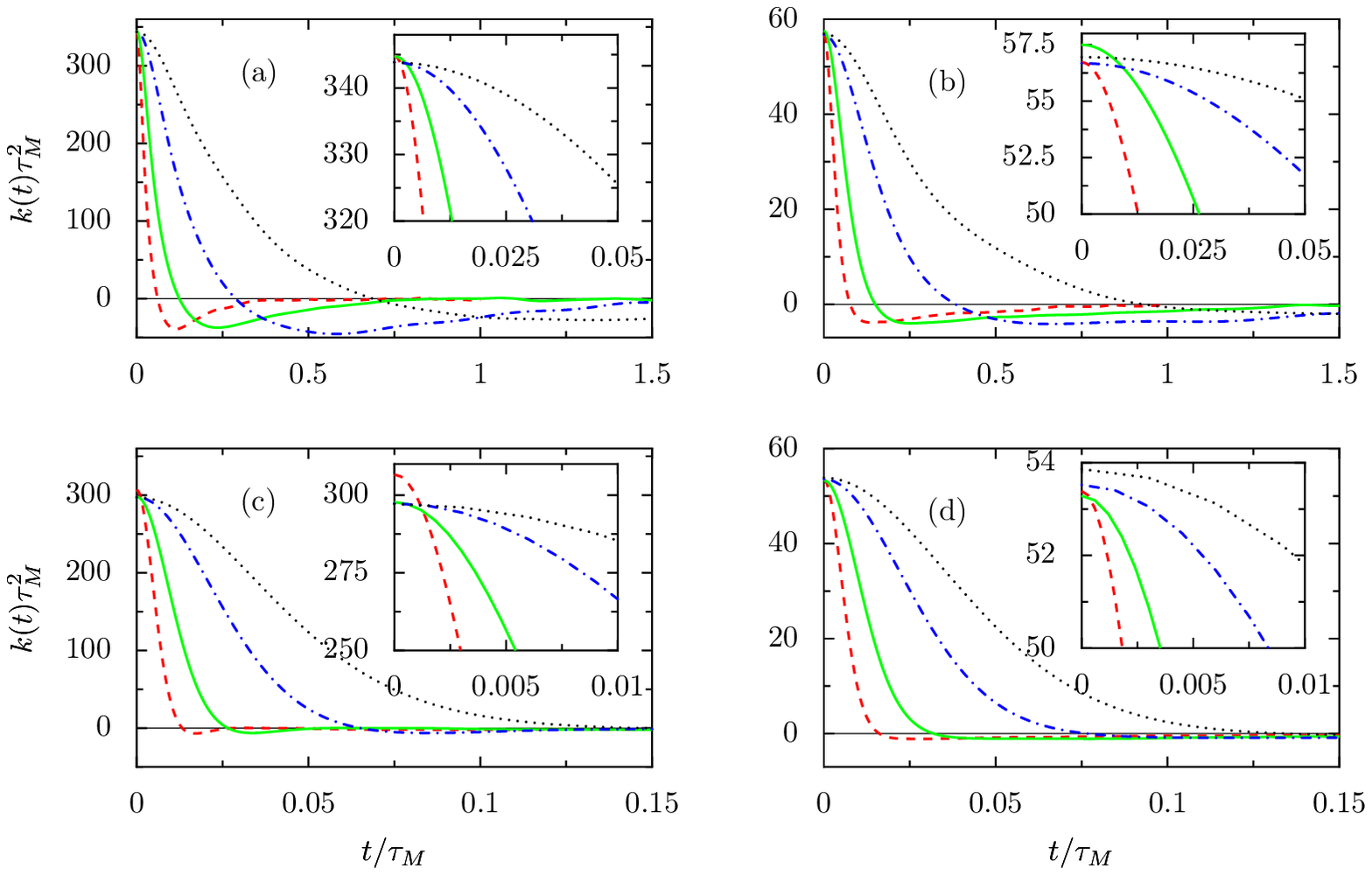}
\caption{(color online) The rescaled memory function $k(t) \tau^2_M$ is displayed in dependence of the macroscopic time $t/\t_M$ for large Brownian particles with diameter $D=10d$ in panels (a) and (b) and small Brownian particles with $D=d$ in panels (c) and (d). Panels (a) and (c) correspond to the fluid density $n^*=0.8$ and panels (b) and (d) to $n^*=0.4$. All panels contain results for different masses $M=100m$ (red, dashed line), $M= 25 m$ (green, solid line), $M=4m$  (blue dash-dotted line), and $M=m$  (black, dotted line). In the insets the short time behavior of the rescaled memory kernel is shown. Note that the rescaled memory function assumes a value at $t=0$ that is almost independent of the mass. This confirms the assumption that the second derivative of the momentum auto-correlation function only weakly depends on the mass. The decay of the memory kernel on the macroscopic time scale though becomes faster with growing mass and also with decreasing diameter, see also Fig.~\ref{f4}. }
\label{f5}
\end{figure*}

As it is obvious from Fig.~\ref{f4} the initial value of the memory kernel $k(0)$ strongly depends on the mass $M$ of the Brownian particle. From the equation of motion of the momentum auto-correlation function, eq.~(\ref{dC}), it follows upon differentiation with respect to time that 
\be
k(0) = -\frac{\ddot{C}(0)}{C(0)}.
\ee{k0}
Assuming that $\ddot{C}(0)$ only weakly depends on mass we obtain with the equipartition law $C(0)= M k_B T$ that $k(0)$ is inversely proportional to the mass of the Brownian particle. For a comparison of the memory kernel at different masses $M$ we therefore scale its value by the mass ratio $m/M$, i.e., we consider the dimensionless reduced memory kernel $k(t)\t_M^2$ as a function of $t/\t_M$ where $\t_M =\t \sqrt{M/m}$ is the relevant, ``macroscopic'' time scale of a Brownian particle of mass $M$. Fig.~\ref{f5} clearly indicates that the rescaled memory kernel at zero time is almost independent of mass. 

The decay of the memory kernel is qualitatively the same for all values of Brownian particle mass and diameter as well as for both fluid densities. It is much faster than the decay of the corresponding momentum auto-correlation functions and for most parameter values consists in a rapid initial decay that overshoots to negative values and then slowly approaches zero from the negative side. In particular, the memory kernel does not display pronounced oscillations in contrast to the momentum auto-correlation function. As for the momentum auto-correlation function our data are not sufficient to specify a particular decay law at large times.

The overall decay is faster both for heavier and smaller particles. Hence, with increasing mass of the Brownian particle an approximation of the memory kernel by a delta function describing instantaneous friction becomes more reliable.  The larger density leads to a larger reduced memory function but hardly influences its shape.

\section{Brownian mass and diameter dependence of the fluctuating force}\label{VII}
\begin{figure}[t]%[tp]
\includegraphics[width=7cm]{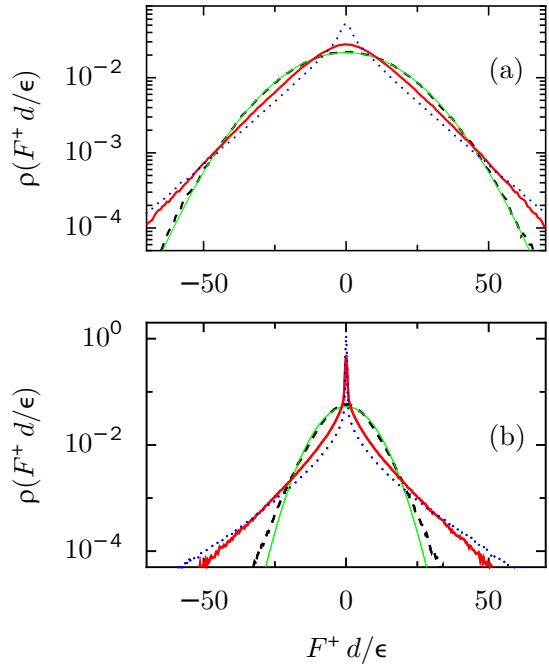}
\caption{(color online) Histograms of the fluctuating forces are displayed for  Brownian particles with mass $M=100m$ and different diameters $D= d$ (blue, dotted line) $D=5d$ (red, solid line) and $D=10 d$ (black broken line). In the latter case a Gaussian distribution with the same mean value and variance is shown for comparison (green, thin solid line). It perfectly coincides with the histogram for the high fluid density $n^*=0.8$ in panel (a) whereas deviations are visible in the tails of the distribution for the lower density $n^*=0.4$ in panel (b).     
}
\label{f6}
\end{figure}
The statistical properties of the fluctuating force strongly depend on the diameter of the Brownian particles and the fluid density.
For small particles and low densities the distributions are leptokurtic with approximately exponential, or even more pronounced tails. For larger Brownian particles and higher fluid densities the distributions approach a more Gaussian shape, see Fig.~\ref{f6}. A distinct dependence of the fluctuating force distribution on the mass of the Brownian particle only exists for small Brownian particles and low fluid densities while it is insignificant otherwise.  
\begin{figure}[t]%[tp]
\includegraphics[width=7cm]{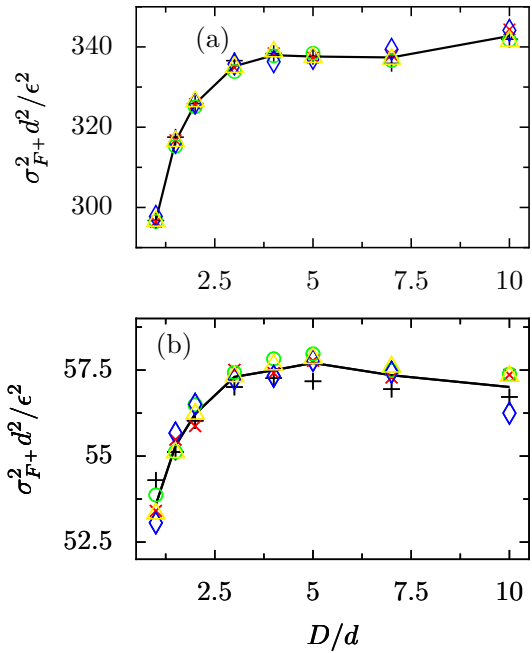}
\caption{(color online) The variance of the fluctuating forces is displayed 
for fluids with densities $n^*=0.8$ and $n^*=0.4$ in the panels (a) and (b), respectively, as functions of the diameter of the Brownian particle $D$ for different masses $M=m$ (black plus), $M=4m$ (red cross), $M=25m$ (green circle), $M=49 m$ (blue diamond)  and $M=100m$ (yellow triangle). The variances for the different masses differ only insignificantly from the behavior  obtained for the average over the masses (black solid line). The variance of the fluctuating forces for the denser fluid is larger, but otherwise the increase at small diameters and the saturation to an approximately constant large particle value agrees qualitatively. 
}
\label{f7}
\end{figure}
\begin{figure}[t]%[tp]
\includegraphics[width=7cm]{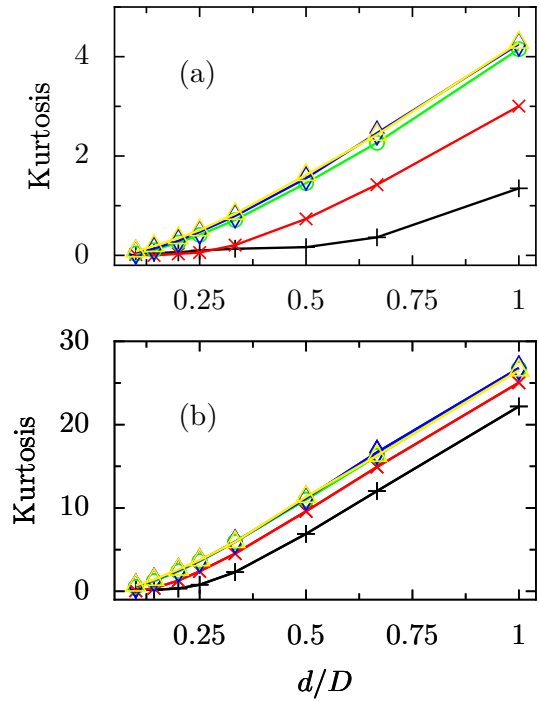}
\caption{(color online) The kurtosis $\langle F^{+4}\rangle/\langle F^{+2}\rangle^2 -3$ of the fluctuating forces is displayed 
for fluids with densities $n^*=0.8$ and $n^*=0.4$ in the panels (a) and (b), respectively, as functions of the inverse diameter of the Brownian particle $1/D$ for different masses $M=m$ (black plus), $M=4m$ (red cross), $M=25m$ (green circle), $M=49 m$ (blue diamond)  and $M=100m$ (yellow triangle). As a guide to the eye, points are connected by solid lines of same color.  For increasing particle size the kurtosis approaches zero. 
}
\label{f8}
\end{figure}
\begin{figure}[t]%[tp]
\includegraphics[width=7cm]{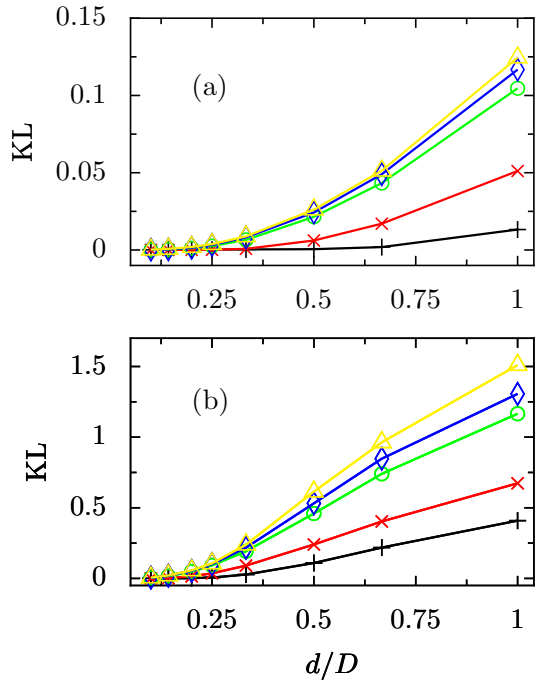}
\caption{(color online) Minimal Kullback-Leibler divergence of the fluctuating force distribution to a Gaussian distribution $KL= \min_{\text{Gaussian} \r^*} KL(\r_{F^+}||\r^*)$ is displayed 
for fluids with densities $n^*=0.8$ and $n^*=0.4$ in the panels (a) and (b), respectively, as functions of the inverse diameter of the Brownian particle $1/D$ for different masses $M=m$ (black plus), $M=4m$ (red cross), $M=25m$ (green circle), $M=49 m$ (blue diamond)  and $M=100m$ (yellow triangle). As a guide to the eye, points are connected by solid lines of same color.  For increasing particle size the $KL$ approaches zero, and hence the distribution of the fluctuating force becomes Gaussian. 
}
\label{f9}
\end{figure}
\begin{figure}[t]%[tp]
\includegraphics[width=7cm]{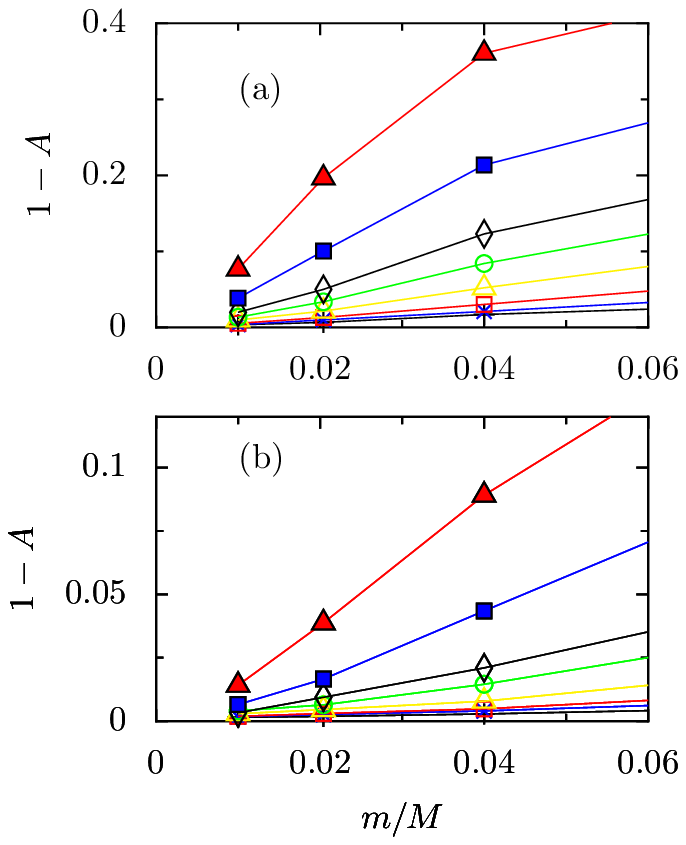}
\caption{(color online) The deviation of the linear regression coefficient from unity, $1-A$ is displayed as a function of inverse mass $1/M$ for fluids with densities $n^*=0.8$ and $n^*=0.4$ in panels (a) and (b), respectively and for Brownian particle diameters $D=d$ (black plus),  $D=1.5d$ (blue cross), $D=2d$ (red open square), $D=3d$ (yellow open triangle), $D=4d$ (green open circle), $D=5d$ (black open diamond), $D=7d$ (blue filled square), and $D=10d$ (red filled triangle). For the sake of better readability the data points are connected by lines with the color of the corresponding symbol. In all cases the approach of $A$ to unity for large values of mass $M$ is approximately linear. It is faster for smaller particles and for the smaller fluid density.   
}
\label{f10}
\end{figure}

The observed statistics of the fluctuating forces can be qualitatively understood in terms of the number of fluid particles that simultaneously interact with the Brownian particle. This number, $N_I$, fluctuates as fluid particles constantly enter and leave the interaction region of the Brownian particle. If, for example, at a given instant of time, no fluid particle happens to interact with the Brownian particle, i.e. if $N_I=0$,  then there will be no force exerted on the particle. This situation may happen more frequently if the fluid density is low and the Brownian particle diameter and therefore also its interaction region is small. This explains the pronounced peak of the fluctuating force distribution at zero force for low fluid density and  small Brownian particles. On the other hand, if the number of fluid particles interacting with the Brownian particle is large, as it is the case for large densities and large Brownian particles, not only the probability of zero force decreases but also very large forces resulting from single particle impacts are likely to be almost compensated by the influence of the other interacting particles and hence the frequency of extremely large forces becomes suppressed. In this way probability is transferred both from the tails and the center towards the flanks of the force distribution when the fluid density or the particle diameter increase. Since the forces that are simultaneously acted on the Brownian particle by different fluid particles will only show weak correlations, the sum of these forces will converge to a Gaussian random number if, on average, the number $N_I$ is sufficiently large. Although these arguments primarily apply to the total force $F(t)$ acting on the Brownian particle it still reflects the observed behavior of the fluctuating force statistics.

In order to characterize the parameter dependence of the fluctuating force statistics in more quantitative terms we estimated the first four moments of the fluctuating force for various parameter vales. All odd moments do not significantly differ from zero. The second moment therefore coincides with the variance of the fluctuating force. It increases with the diameter $D$ and approaches a constant value for large diameters. At lower fluid densities the variance is smaller but it hardly depends on the mass $M$ of the Brownian particle. These features are illustrated in   
Fig.~\ref{f7}.

From the second and fourth moment of the fluctuating force the kurtosis is determined as $\langle F^{+4} \rangle/\langle F^{+2} \rangle^2 -3$. As it measures the excess of probability located in the center and the tails of a distribution as compared to a Gaussian distribution, it is larger for the less dense fluid and approaches zero with increasing diameter of the Brownian particle, see Fig.~\ref{f8}, in agreement with the qualitative arguments discussed above.    

Whereas the kurtosis only specifies a particular aspect of a deviation from a Gaussian distribution, the Kullback-Leibler divergence \cite{K} allows one to quantify a distance measure of a given distribution from the closest Gaussian distribution. According to its definition the Kullback-Leibler divergence of $\r^*$ to $\r$ is given by 
\be
KL(\r||\r^*) = \i d x \r(x) \ln\frac{\r}{\r^*}
\ee{KL}
where $\r$ and $\r^*$ denote probability density functions. The Kullback-Leibler divergence in general is positive. It vanishes if and only if the probability density functions $\r$ and $\r^*$ coincide with each other \cite{K}. It is straightforward to show that the Gaussian distribution $\r^*$  with the mean value and variance of a distribution $\r$ has the minimal divergence from $\r$ compared to all other Gaussian distributions. The minimal Kullback-Leibler divergence of a Gaussian to the fluctuating force distribution is displayed in Fig.~\ref{f9}. In particular, it corroborates that the fluctuating force indeed approaches a Gaussian distribution with increasing diameter of the Brownian particle.    

Finally we investigated the relation between the total and the fluctuating forces $F(t)$ and $F^+(t)$. We first discuss the limit of large masses of the Brownian particle. Since the memory kernel decreases with increasing mass $M$ whereas the fluctuating force correlation approaches a finite function in this limit, the organized contribution to the total force becomes negligible compared to the fluctuating force. Therefore one expects that the total and the fluctuating force agree with each other for Brownian particles with sufficiently large mass compared to the fluid particle mass. In the literature 
\cite{BZ} one finds as a rough estimate of the deviations 
\be
F^+(t) = (1+ O((m/M)^{1/2})) F(t)
\ee{FpFt}
To investigate the relation between $F^+(t)$ and $F(t)$ in more detail we determined the distributions of the fluctuating force conditioned on the total force within intervals extending over one force unit. Apart from Brownian particles that are small and light at the same time it turned out that the distributions of the conditional fluctuating forces are independent of the conditioning value of $F(t)$ except for an overall shift which is linear in the condition. Therefore the relation between the fluctuating and the total force can be described by the model 
\be
F^+_F = A F + \xi
\ee{FpAFt}
where $F^+_F$ denotes the fluctuating force $F^+$ conditioned on the total force restricted to the interval $[F-0.5\e/d,F+0.5\e/d)$. Here, the residue $\xi$ is independent of the condition $F$ and  Gaussian distributed. Values of the parameter $A$ were estimated by means of linear regression based on eq. (\ref{FpAFt}). Deviations of $A$ from unity are displayed in Fig.~\ref{f10} as functions of inverse mass. Apparently $A$ approaches unity linearly in $1/M$ as the mass goes to infinity. So actually the convergence of the fluctuating force towards the total force is found to be faster than indicated by eq.~(\ref{FpFt}).

\section{Conclusions}\label{VIII}
We presented an extensive numerical study of the motion of single Brownian particles with various masses and diameters interacting with   fluid  particles. The data obtained from the MD simulations were analyzed in terms of a generalized Langevin equation of Mori type. The memory kernel was estimated on the basis of the total force-momentum correlation function and the total force auto-correlation functions both of which can directly be obtained from the MD simulation data. The consistency of the obtained memory kernel and the fluctuating forces was tested in terms of the fluctuation dissipation theorem and an identity for the fluctuating force momentum correlation function. We discussed the momentum auto-correlation function for different Brownian particle masses and diameters and found pronounced oscillations for large and light particles. The maximal value of the memory function is inversely proportional to mass. Its decay on the macroscopic time scale $\t_M=\t \sqrt{M/m}$ is faster for smaller and/or heavier particles. This is in qualitative agreement with Ref. \cite{HML} saying that the Markovian limit of Brownian dynamics is approached if the ratio of fluid and mass {\it densities} approaches zero.      

The distribution of the fluctuating forces is leptokurtic for small particles and approaches a Gaussian form for larger particles. The approach to a Gaussian is faster for light particles. For Brownian particles being ten times larger than the fluid particles the fluctuating force was found to be perfectly Gaussian for all considered masses up to 100 times the fluid particle mass. This observation is based on direct comparison of histograms, on estimates of the kurtosis and on the Kullback-Leibler divergence of the closest Gaussian. 

For massive Brownian particles the difference between the fluctuating and the total force was shown to shrink proportionally to the mass ratio $m/M$.

%---------------------------------------------------------------------
\section*{Acknowledgments}

This paper is dedicated to Professor Peter H\"anggi, on occasion of his $60^\textrm {th}$ birthday. We wish him many scientifically fertile years still to come. This work was supported by the German Excellence Initiative via the \textit {Nanosystems Initiative Munich}(NIM), by the Deutsche Forschungsgemeinschaft under the Grant HA 1517/25-2, and by Basic Science Research Program through the National Research Foundation of Korea (NRF) funded by the Ministry of Education, Science and Technology (Grant number 2010-0013812).

\appendix
\section{Proof of eq. (\ref{FpP}) and the fluctuation dissipation theorem}\label{A}
We first show that eq.~(\ref{ssLE}) in combination with eq.~(\ref{dC}) implies eq. (\ref{FpP}). Solving the stationary generalized Langevin equation (\ref{ssLE}) for the fluctuating force yields
\be
F^+(t) = F(t) + \i_{-\infty}^t ds\: k(t-s) P(s)\:,
\ee{A1}
where the total force $F(t)$ gives rise to the acceleration of the Brownian particle. i.e.
\be
\dot{P}(t) = F(t)\:.
\ee{A2}
By multiplying both sides of eq. (\ref{A1}) with $P(0)$ and performing an equilibrium average we obtain
\be
\begin{split}
&\langle F^+(t) P \rangle =   \langle \dot{P}(t) P(0) \rangle +\!
\i_{-\infty}^t\! \!ds \: k(t\!-\!s) \langle P(s) P(0) \rangle\\
&= \dot{C}(t) + \!\i_0^t \!ds\: k(t\!-\!s) C(s)
 + \!\i_{-\infty}^0 \!ds \:  k(t\!-\!s)
C(s)\:.
\end{split}
\ee{A3}
The first two terms of the right hand side of the last line cancel each other as a consequence of  eq. (\ref{dC}) such that only the third term remains. Changing in this remaining term the variable of integration $s \to -s$ and observing the symmetry of the correlation function $\langle P(t) P(0) \rangle \equiv C(t) = C(-t)$ one obtains the expression on the right hand side of eq. (\ref{FpP}). This proves eq. (\ref{FpP}).

In order to prove that the fluctuation dissipation theorem follows from the generalized Langevin equation (\ref{ssLE}) in combination with the form of the fluctuating force-momentum correlation eq. (\ref{FpP}) and the equation of motion of the momentum auto-correlation function eq. (\ref{dC})  we determine the auto-correlation function of the fluctuating force using eq. (\ref{A1}). We then obtain with eq. (\ref{A1})
\be
\begin{split}
&\langle F^+(t) F^+(0) \rangle = \langle F^+(t) F(0) \rangle\\
&+\i_{-\infty}^0 ds \: k(-s) \langle F^+(t) P(s) \rangle\\
&=  - \ddot{C}(t)
- \i_{-\infty}^t ds\: \:k(t-s)
\dot{C}(s) \\
& +\i_{-\infty}^0 ds\:
 \i_0^\infty ds'\: k(-s) k(t-\!s+\!s') C(s')\\
&=k(t) C(0) + \i_0^\infty ds\: k(t+s) \dot{C}(s)\\
&+ \i_0^\infty ds\i_0^\infty ds'\: k(s) k(t+s+s') C(s')\:,
\end{split}
\ee{A4}
where we used eq. (\ref{FpP}) at the second equal sign and eq. (\ref{ddC}) at the third equal sign. Using eq. (\ref{dC}) one may show with little algebra that the last two terms compensate each other such that the fluctuation dissipation theorem is recovered.

Vice versa, if one takes for granted the stationary generalized Langevin equation (\ref{ssLE}) together with the fluctuation dissipation theorem (\ref{sfdt}) and the fluctuating force-momentum correlation function (\ref{FpP}), one finds for the momentum auto-correlation function the following equation of motion
\be
\begin{split}
\dot{C}(t)& = - \i_{-\infty}^t ds \:k(t-s) C(s) + \langle F^+(t) P(0) \rangle \\
&=  - \i_0^t ds \:k(t-s) C(s)\:,
\end{split}
\ee{A5}
where we split the integral on the right hand side of the first line into a part extending from $-\infty$ to $0$ and another one from $0$ to $t$. The first contribution is exactly canceled by the fluctuating force-momentum correlation given by eq. (\ref{FpP}). Hence we recover eq. (\ref{dC}) as the equation of motion for the momentum auto-correlation function.

%\section{Conclusions}

\end{document}